\documentclass[a4paper,fleqn,usenatbib]{mnras}

\usepackage{newtxtext,newtxmath}
 
\usepackage{enumitem}

\usepackage[T1]{fontenc}
\usepackage{ae,aecompl}

\usepackage{graphicx}	
\usepackage{amsmath}	
\usepackage{amssymb}	

\defcitealias{1975ApJ...195..157C}{CAK}
\defcitealias{1984ApJ...284..337O}{OR84}
\defcitealias{1985ApJ...299..265O}{OR85}
\defcitealias{1990ApJ...349..274R}{ROC90}

\title[Alfv\'en wave damping in magnetic line-driven winds]{Line-drag damping of Alfv\'en waves in radiatively driven winds of magnetic massive stars}

\author[F.~A.~Driessen, N.~D.~Kee, J.~O.~Sundqvist, and S.~P.~Owocki]{
F.~A.~Driessen$^{1}$\thanks{E-mail: florian.driessen@kuleuven.be},
N.~D.~Kee$^{1}$,
J.~O.~Sundqvist$^{1}$, and 
S.~P.~Owocki$^{2}$
\\
$^{1}$Institute of Astronomy, KU Leuven, Celestijnenlaan 200D, 3001 Leuven, Belgium\\
$^{2}$Bartol Research Institute, Department of Physics \& Astronomy, University of Delaware, Newark, DE 19716, USA\\
}

\date{Accepted XXX. Received YYY; in original form ZZZ}

\pubyear{2020}

\begin{document}
\label{firstpage}
\pagerange{\pageref{firstpage}--\pageref{lastpage}}
\maketitle

\begin{abstract}
Line-driven stellar winds from massive (OB) stars are subject to a strong line-deshadowing instability. Recently, spectropolarimetric surveys have collected ample evidence that a subset of Galactic massive stars hosts strong surface magnetic fields. We investigate here the propagation and stability of magneto-radiative waves in such a magnetised, line-driven wind. Our analytic, linear stability analysis includes line-scattering from the stellar radiation, and accounts for both radial and non-radial perturbations. We establish a bridging law for arbitrary perturbation wavelength after which we analyse separately the long- and short-wavelength limits. While long-wavelength radiative and magnetic waves are found to be completely decoupled, a key result is that short-wavelength, radially propagating Alfv\'en waves couple to the scattered radiation field and are strongly damped due to the line-drag effect. This damping of magnetic waves in a scattering-line-driven flow could have important effects on regulating the non-linear wind dynamics, and so might also have strong influence on observational diagnostics of the wind structure and clumping of magnetic line-driven winds.
\end{abstract}

\begin{keywords}
radiative transfer -- waves -- instabilities -- stars: early-type -- stars: magnetic field -- stars: winds, outflows
\end{keywords}



\section{Introduction}

The powerful radiation from massive (OB) stars is able to transfer its momentum to the stellar wind plasma by absorption and scattering in spectral lines. A quantitative description of this line-driving has been provided by the seminal work of \citet*[][hereafter CAK]{1975ApJ...195..157C}. In formulating this nowadays widely used theory, \citetalias{1975ApJ...195..157C} relied on the Sobolev approximation for describing the radiative acceleration of the wind. This approximation assumes that in a highly supersonic outflow the spectral line transport can be described based on purely \emph{local} information \citep{1960mes..book.....S} meaning that the hydrodynamic flow quantities are constant over a Sobolev length $L_\mathrm{Sob} \equiv \varv_\mathrm{th}/(d\varv_n /dn)$ (with $\varv_\mathrm{th}$ the ion thermal speed and $d\varv_n /dn$ the projected velocity gradient in radiation direction $n$).

It turns out that the line-driving mechanism is subject to a radiative instability as first suggested by \citet{1970ApJ...159..879L}. This very strong, intrinsic instability is known as the \emph{line-deshadowing instability} (LDI). \citet*{1979ApJ...231..514M} and \citet{1980ApJ...241.1131C} performed a linear stability analysis under the assumption of optically thin perturbations finding an unstable wind due to the LDI. On the other hand, however, \citet{1980ApJ...242.1183A} assumed that perturbations follow the Sobolev approximation and found that the instability vanishes. Instead, a stable, inward propagating wave (Abbott wave) arises. 

These contradictory results were reconciled by \citet[][hereafter OR84]{1984ApJ...284..337O} showing that both cases occur depending on perturbation wavelength, i.e.~short-wavelength (instability) and long-wavelength (waves). Their analysis illustrates that the LDI is inherently acting on spatial scales below or near the Sobolev length. This naturally explains why the LDI does not occur in dynamical \citetalias{1975ApJ...195..157C} models that rely on the Sobolev approximation to compute the radiative acceleration. Later analytic work performed by Owocki \& Rybicki extended these results to clarify the effects of line-scattering on the growth rate of the LDI \citep[][hereafter OR85]{1985ApJ...299..265O}, what the spatial and temporal evolution of the LDI is \citep{1986ApJ...309..127O}, the effect of non-radial perturbations on the growth rates \citep*[][hereafter ROC90]{1990ApJ...349..274R}, and the growth rates in flows that have an optically thick continuum \citep{1991ApJ...368..261O}. 

A calculation of \citetalias{1984ApJ...284..337O} shows that the LDI quickly attains a non-linear regime, having a typical growth rate on the order of $e^{100}$ (including line-scattering this becomes $e^{50}$) in a typical O-star wind. Subsequent numerical simulations of the non-linear evolution of the LDI \citep*[e.g.][]{1988ApJ...335..914O,1999ApJ...510..355O,2017MNRAS.469.3102F,2018A&A...611A..17S,2019A&A...631A.172D} have shown that the wind fragments and forms large-scale slow, overdense, \emph{clumpy} structures that are separated by a fast, nearly void medium and is quite different from the homogeneous winds predicted by \citetalias{1975ApJ...195..157C}. 

The cause of large-scale structures seen in non-linear numerical simulations of the LDI is complex. The initial instability is due to short-wavelength perturbations as predicted by linear analysis. Recently, \citet{2017MNRAS.469.3102F} suggested that in the accelerating wind additionally these wave perturbations get coherently stretched over many Sobolev lengths thereby allowing the non-linear growth of the LDI to proceed to high amplitudes. This overall process would lead then to the typical wind structure of clumps separated by a nearly void medium in numerical simulations.

Observations of line-driven winds from massive stars have indeed provided evidence of clumpy structures. For example, in non-magnetic massive star winds it is well known that the LDI leads to significant \emph{wind clumping}, with important effects on observational diagnostics \citep*[e.g.,][for an overview]{2015IAUS..307...25P}.  This has lead to a basic understanding of observed phenomena such as soft X-ray emission (\citealt{1997A&A...322..167B}, \citealt*{1997A&A...322..878F}), extended regions of zero residual flux in resonance UV spectral lines \citep*{1983ApJ...274..372L,2012ASPC..465..119S}, and migrating subpeaks in optical recombination spectral lines \citep*{1998ApJ...494..799E,2002A&A...383.1113D}. 

It is important to recognize that the above theoretical and observational analyses have all been carried out for non-magnetic massive star winds. Yet, two decades ago spectropolarimetric observations have made clear that some massive stars in our Galaxy are magnetic \citep{2002MNRAS.333...55D}. Over the years dedicated spectropolarimetric surveys of massive stars in our Galaxy (\citealt{2015A&A...582A..45F} (BOB); \citealt{2016MNRAS.456....2W} (MiMeS)) have gathered ample observational evidence that a modest fraction ($\sim$10\%) of massive stars harbour strong ($\sim$kG), organised (often dipolar) surface magnetic fields. The occurrence of such surface magnetic fields is none the less puzzling because massive stars lack any subsurface hydrogen/helium recombination zones that are thought to generate dynamo magnetic fields as, for example, is the case for the Sun. The accepted view is that the surface magnetic fields are of primordial origin, but the physical mechanism that can explain the generation and incidence of these fields is not yet settled \citep[e.g., see][for recent insights]{2019Natur.574..211S}. 

Previous theoretical work on line-driven winds from magnetic massive stars studied already the dynamics in various settings \citep{2002ApJ...576..413U,2006ApJ...640L.191U,2008MNRAS.385...97U}. A major result from these studies is that the stellar wind is channeled along the magnetic field lines forming a so-called \emph{circumstellar magnetosphere} that can significantly alter the mass-loss rate of the stars. However, all these theoretical works applied CAK theory to study the global wind dynamics in these magnetic environments. As discussed above, this means the dynamics of the LDI is neglected in those models. Reasons of omission have largely been due to the significant computational complexity of non-linear LDI simulations already for non-magnetic line-driven winds. With the increase in detections of magnetic massive stars, however, it has become important to undertake an investigation of the interplay between the LDI and stellar magnetic fields.

From a theoretical point of view several questions arise on what the influence is of the magnetic field on the development of the instability. In environments with sufficiently strong magnetic fields the fluid flow will be confined to the magnetic field lines. Such physics might severely constrain, or even prevent, any horizontal fragmentation of matter, thereby reducing the clumpiness of the wind. If true, this would imply that the typical structure of dense, clumps separated by a nearly void medium might be replaced by large, dense shells of matter separated by a nearly void medium. Another specific question to address is how the additional waves associated with the magnetic field behave in the wind: What are their propagation properties? How do they couple to the radiation? And, most importantly, do they amplify or attenuate any horizontal fragmentation of matter?

As a first step, in the present paper we perform an analysis of linear perturbations in a magnetic line-driven wind. An initial understanding of the magneto-radiative wave propagation and instability growth can be gained in the \emph{linear} regime using analytical tools. This serves to interpret future \emph{non-linear} magnetic numerical simulations of the LDI.

\section{Theoretical formulation}

\subsection{Assumptions}\label{sec:assump}

To reduce the complexity of the problem we adopt a set of assumptions to make calculations tractable while still retaining the core physics. This is done both for the stellar wind plasma and the radiation field by treating a local volume of wind plasma above magnetic pole of a stellar dipole field. 

We take standard OB star wind assumptions for the background, mean wind and neglect gas pressure gradient and continuum opacities. The mean magnetic field above the pole is taken to be purely radial while radiation from a point star is radially streaming and interacts with one isolated pure scattering line for linear stability. The line-scattering is taken to be isotropic with complete frequency redistribution. All perturbations are linear and act on WKB order, i.e.~perturbation wavelengths are much smaller than the typical spatial scales of the wind.

%
%

Recall that in a magnetic flow the basic modes are due to magnetic tension (Alfv\'en mode) and magnetic pressure (fast and slow mode) from the magnetic field line. In the absence of thermal pressure (zero-sound-speed limit), however, the slow mode vanishes while the fast mode propagates isotropically at the Alfv\'en speed. Therefore, what remains in our present problem are radiative, Alfv\'en, and fast modes.

Our assumption of a point star and the omission of thermal pressure have additional consequences for the nature of the flow critical points \citep{1980ApJ...242.1183A}. However, we do not analyse the solution topology in this work, and as such here these assumptions are relatively minor.

\subsection{Magnetohydrodynamics of line-driven winds}\label{sec:mhdpart}

We first present a standard linear perturbation analysis for the magnetic flow in the presence of a radiative force. Readers familiar with this may wish to proceed to \S \ref{sec:radfieldpart} where we discuss the essential background on the radiation field.

\subsubsection{Governing equations}

We adopt the equations of ideal magnetohydrodynamics (MHD) (appropriate for a highly ionised medium like a line-driven wind) to describe the magnetic wind dynamics. The energy equation is omitted by assuming an isothermal wind. Since thermal pressure is neglected in the present paper no pressure dependencies arise. Furthermore, we describe line-driving based on the radiative acceleration due to a single line and include gravity
\begin{equation}\label{eq:first_mhd_equation}
\frac{\partial \rho}{\partial t} + \rho \frac{\partial \varv_i}{\partial r_i} + \varv_i \frac{\partial \rho}{\partial r_i}= 0,
\end{equation}
\begin{equation}
\rho \left( \frac{\partial \varv_i}{\partial t} + \varv_j \frac{\partial \varv_i}{\partial r_j} \right) = -\rho \frac{GM_{\mathrm{eff}} r_i}{r^3} +  \rho g_i + \frac{1}{4\pi} \epsilon_{ijk}\epsilon_{jlm} \frac{\partial B_m}{\partial r_l} B_k,
\end{equation}
\begin{equation}
\frac{\partial B_i}{\partial t} = \epsilon_{ijk}\epsilon_{klm} \frac{\partial}{\partial r_j}(\varv_l B_m),
\end{equation}
\begin{equation}\label{eq:last_mhd_equation}
\frac{\partial B_i}{\partial r_i} = 0.
\end{equation}
with $\rho$ the density, $\varv_i$ the gas velocity, and $B_i$ the magnetic field (spatial coordinate indices $i$ run from $1\rightarrow 3$). Furthermore, $G$ is the gravitational constant, $M_\mathrm{eff} = (1-\Gamma_e)M_\star$ the effective stellar mass reduced by the effect of (constant) electron scattering in the wind described by Eddington's gamma $\Gamma_e = \kappa_e L_\star/(4\pi G M_\star c)$ with electron scattering opacity $\kappa_e = 0.34$ cm$^2$ g$^{-1}$, and $\epsilon$ is the three-dimensional Levi--Civita tensor. All fluid quantities are assumed to be a function of position $r_i$ and time $t$. In the momentum equation the source terms are gravity, the radiative acceleration due to lines $g_i$, and the Lorentz force from the magnetic field, respectively. The only unknown term here is $g_i$ for which one could use, e.g. the Sobolev approximation for the mean flow, but as alluded in the Introduction this is not valid on the small spatial scales where the LDI operates. Determining this source term for linear perturbations is therefore important, and is discussed in \S \ref{sec:radfieldpart}.

\subsubsection{Linear perturbations in three dimensions}

We apply small (linear) perturbations on the above ideal MHD equations. We write the variables such that they consist of an unperturbed, steady-state part (subscript 0) and a perturbed, evolving part (denoted by a leading $\delta$). By definition a perturbed quantity $\delta q$ has to satisfy $|\delta q|/q_0 \ll 1$. This leads to the following definitions
\begin{equation}
\begin{aligned}
&\rho \equiv \rho_0(r_j) + \delta \rho(r_j,t), \qquad \varv_i \equiv \varv_{i0}(r_j) + \delta \varv_i(r_j,t), \\
&B_i \equiv B_{i0}(r_j) + \delta B_i(r_j,t), \qquad g_i \equiv g_{i0}(r_j) + \delta g_i(r_j,t).
\end{aligned}
\end{equation}

From our local, WKB order analysis density stratifications are understood to be small such that perturbations on gravity can be neglected (this is valid if perturbations are much smaller than the density scale height). Substituting these expressions into \eqref{eq:first_mhd_equation} -- \eqref{eq:last_mhd_equation} and keeping terms up to first order gives on WKB order the MHD perturbed state
\begin{equation}\label{eq:first_mhd_perturbed}
\frac{\partial \delta \rho}{\partial t} + \rho_0 \frac{\partial \delta \varv_i}{\partial r_i} = 0
\end{equation}
\begin{equation}
\rho_0 \frac{\partial \delta \varv_i}{\partial t} = \rho_0 \delta g_i + \frac{1}{4\pi} \epsilon_{ijk}\epsilon_{jlm}\frac{\partial \delta B_m}{\partial r_l}B_{k0},
\end{equation}
\begin{equation}
\frac{\partial \delta B_i}{\partial t} = \epsilon_{ijk}\epsilon_{klm} \frac{\partial}{\partial r_j} (\delta \varv_l B_{m0}),
\end{equation}
\begin{equation}\label{eq:last_mhd_perturbed}
\frac{\partial \delta B_i}{\partial r_i} = 0,
\end{equation}
in a frame locally comoving with the flow. On WKB order a local spherical coordinate system is indistinguishable from a local Cartesian coordinate system. Therefore, the remaining analysis of the perturbed state uses Cartesian $x_j$ in place of spherical $r_j$.

The perturbed MHD equations \eqref{eq:first_mhd_perturbed} -- \eqref{eq:last_mhd_perturbed} are linear and have constant coefficients. This implies that on an unbounded domain their solution can be decomposed into plane waves
\begin{equation}
\begin{pmatrix}
\delta \rho(x_j,t) \\
\delta \varv_i(x_j,t)\\
\delta B_i(x_j,t)\\
\delta g_i(x_j,t)\\
\end{pmatrix}
=
\begin{pmatrix}
\delta \tilde{\rho} \\
\delta \tilde{\varv}_i \\
\delta \tilde{B}_i \\
\delta \tilde{g}_i \\
\end{pmatrix}e^{i ( k_j x_j - \omega t)},
\end{equation}
with $\delta \tilde{\rho}$, $\delta \tilde{\varv}$, $\delta \tilde{B}$, and $\delta \tilde{g}$ the real amplitude of the corresponding perturbed variable (from now on the tilde will be dropped). The wave is described by a wave vector $k_j$ and a wave frequency $\omega$. For real $k_j$, $\omega$ can be real, imaginary, or complex: $\omega \equiv \omega_R + i\omega_I$. Under the plane wave Ansatz the real part $\omega_R$ yields the wave propagation speed (phase speed) while the imaginary part $\omega_I$ is the temporal growth rate and signifies wave growth $(\mathrm{instability},\omega_I>0)$ or wave damping $(\omega_I<0)$. 

With plane waves the partial differential equations of the perturbed MHD state  \eqref{eq:first_mhd_perturbed} -- \eqref{eq:last_mhd_perturbed} reduce to a set of algebraic equations
\begin{equation}
-i\omega \delta \rho + i\rho_0 k_i \delta \varv_i = 0,
\end{equation}
\begin{equation}
-i\omega \rho_0 \delta \varv_i = \rho_0 \delta g_i + i\frac{1}{4\pi} ( B_{j0} k_j \delta B_i - B_{j0} \delta B_j k_i  ),
\end{equation}
\begin{equation}
-i\omega \delta B_i = i\left( k_j B_{j0} \delta \varv_i - k_j \delta \varv_j B_{i0} \right),
\end{equation}
\begin{equation}
i k_i \delta B_i = 0,
\end{equation}
where we have used the relation $\epsilon_{ijk}\epsilon_{klm} = \delta_{il} \delta_{jm} - \delta_{im} \delta_{jl}$ to express the Levi--Civita tensor in terms of the Kronecker delta.

The 1st equation (perturbed continuity) is decoupled from the other equations. Likewise the 4th equation (perturbed divergence-free condition) provides a constraint on $\delta B_i$. The latter reads that waves propagating along the mean magnetic field $B_i$ will result in a perturbed magnetic field perpendicular to the wave propagation. The 2nd (momentum) and 3rd (induction) equation are coupled to each other. To investigate the stability properties of the flow it is beneficial to work in a perturbed velocity representation. Additionally, this naturally expresses what the linear response of the radiative acceleration is due to a velocity perturbation, i.e.~it accounts for Doppler shifts,
\begin{equation}\label{eq:MHDdispersionrelation}
\begin{aligned}
\Biggl\{ &\left[ \omega^2 - \frac{1}{4\pi \rho_0}(k_l B_{l0})^2 \right] \delta_{ij} - i\omega \frac{\delta g_i}{\delta \varv_j}  \\
&+ \frac{1}{4\pi \rho_0} \left[ \left( k_l B_{l0} \right) \left( k_i B_{j0} + B_{i0} k_j  \right) - B_0^2 k_i k_j \right]  \Biggr\} \delta \varv_j = 0 ,
\end{aligned}
\end{equation}
or the equivalent eigenproblem
\begin{equation}
\mathcal{M}_{ij} \delta \varv_j = 0.
\end{equation}

Once the velocity perturbation is set, the eigenproblem establishes a dispersion relation between $\omega$ and $k_j$ and allows to study wave propagation and instability growth. In full generality $\mathcal{M}_{ij}$ is a rather involved eigentensor, preventing a consideration of its complete representation. Instead, we here invoke our assumption of a radial mean magnetic field chosen to point along the $x_3$-axis ($z$-axis) of the local Cartesian coordinate system. Without loss of generality, we consider an arbitrary wave propagation in the two-dimensional $x_1$-$x_3$ plane around the radial axis, i.e. $k_1 = k \sin \theta$ and $k_3 = k \cos \theta$, with $\theta$ the angle between the wave vector and the mean magnetic field. Symmetry properties make the physical interpretation in the $x_2$-$x_3$ plane the same. Under these simplifications the eigentensor becomes
\begin{equation}\label{eq:eigentensor}
\begin{aligned}
\mathcal{M}_{11} &= \omega^2 -\varv_A^2 (k_1^2 + k_3^2) - i \omega \frac{\delta g_1}{\delta \varv_1}, \\
\mathcal{M}_{22} &= \omega^2 -\varv_A^2  k_3^2 - i \omega \frac{\delta g_2}{\delta \varv_2}, \\
\mathcal{M}_{33} &= \omega^2 - i \omega \frac{\delta g_3}{\delta \varv_3}, \\
\mathcal{M}_{ij} &= - i \omega \frac{\delta g_i}{\delta \varv_j}, \qquad i\neq j
\end{aligned}
\end{equation}
and the Alfv\'en speed is introduced as $\varv_A = B_0/\sqrt{4\pi \rho_0}$.

\subsection{Response of the radiative acceleration to a velocity perturbation}\label{sec:radfieldpart}

In the dispersion relation \eqref{eq:MHDdispersionrelation} we require the linear response of the radiative acceleration to a velocity perturbation, $\delta g_i/\delta \varv_j$. This tensor expression was first derived in \citetalias{1990ApJ...349..274R} showing that the multi-dimensional nature adds additional subtle radiation effects. As this tensor expressing $\delta g_i/\delta \varv_j$ lies at the core of the present paper, we discuss here the important physical terms it consists of. The reader interested in a more fundamental understanding of its nature and the techniques required to derive is referred to Appendix \ref{sec:appendix}. 

Following \citetalias{1990ApJ...349..274R} the perturbed radiative acceleration tensor for a single line is
\begin{equation}\label{eq:DgDvFinal}
\begin{aligned}
\frac{\delta g_i}{\delta \varv_j} = {} &\frac{g_{0}}{\varv_\mathrm{th} \langle \mu \mathcal{D}_\mu p_\mu \rangle} \\
&\times \left\langle i n_i \left( \mathcal{D}_\mu - (1-s) \frac{\langle \mathcal{D}_\mu p_\mu \rangle}{\langle p_\mu \rangle}  \right)\frac{n_l k_l n_j}{Q_0} \frac{Q_0 / k_L}{ 1 + i n_l k_l/Q_0} \right\rangle,
\end{aligned}
\end{equation}
where the meaning of the variables is introduced in the Appendix. Recalling our assumptions (\S \ref{sec:assump}) that include the point-star approximation, $\mathcal{D}_\mu = \delta(\mu -1)$, and that linear stability of the flow is considered for a single optically thick line, $\tau_\mu \gg 1$, such that the escape probability becomes $p_\mu \approx 1/\tau_\mu = Q_0/k_L$, the prefactor outside angle brackets in Eq.~\eqref{eq:DgDvFinal} reduces to
\begin{equation}
 \frac{g_{0}}{\varv_\mathrm{th} \langle \mu \mathcal{D}_\mu Q_0 \rangle} = \frac{2 \omega_\star}{\chi_\star}.
\end{equation}

We define the growth rate $\omega_\star$, characterizing how fast the instability grows per unit time, as the ratio of the \emph{mean} radiative acceleration of a single line $g_{0}$ to the thermal speed of ions in the wind $\varv_\mathrm{th}$. The quantity $\chi_\star^{-1} = 1/(2 \langle \mu \mathcal{D}_\mu Q_0 \rangle)$ is the bridging length for a single optically thick line  and is comparable to the Sobolev length \citepalias{1984ApJ...284..337O}. It can be interpreted as the typical spatial scale over which a photon gets absorbed at high frequencies within a line profile.

Another important term is contained within the parentheses. The quantity $s$ is the line photon destruction probability for an interaction between a photon and a single line. This photon destruction probability allows pure absorption $(s = 1)$ and pure scattering ($s = 0$) to be treated separately. In the present paper we only take into account pure scattering as line-driven winds are mainly driven by line-scattering processes. This then naturally splits Eq.~\eqref{eq:DgDvFinal} in a contribution from direct radiation $\mathcal{D}_\mu$ and scattered radiation $\langle \mathcal{D}_\mu p_\mu \rangle /\langle p_\mu \rangle$. In particular, the scattered radiation has important physical consequences for radiative instabilities by acting as a drag force. In a purely radial flow the scattering \emph{reduces} the absolute growth of the LDI (\citealt{1984ApJ...284..351L}, \citetalias{1985ApJ...299..265O}). On the other hand, when taking into account lateral radiation, scattering acts to \emph{damp} the LDI in the lateral directions \citepalias{1990ApJ...349..274R}. 

With this lateral damping from line-drag in mind one might expect a priori some consequences in a magnetic wind. For example, Alfv\'en waves propagate as transversal waves along a magnetic field and they might be expected to experience effects related to this damping. Whether line-drag affects the magnetic waves, and if so, in what way, is investigated next.

\section{Stability--dispersion analysis}\label{sec:stabdisp}

\subsection{Dependence on perturbation wavelength: bridging law}\label{sec:bridginglaw}

\begin{figure*}
\centering
	\includegraphics[width=\textwidth]{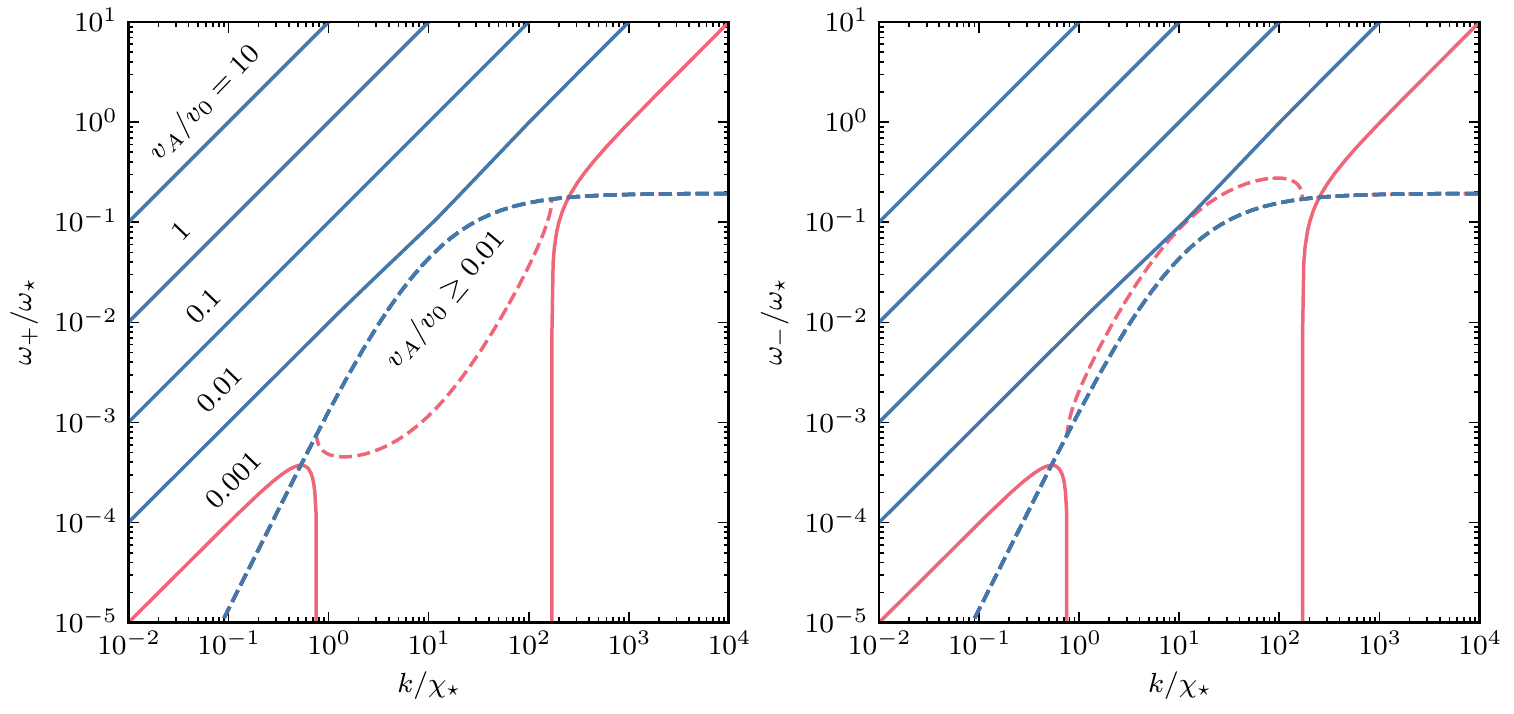}
    \caption{Stability-dispersion solutions for magneto-radiative modes for various values of $\varv_A/\varv_0$ at a point $r=5R_\star$ in the wind. (Left panel) Outward propagating modes $\omega_+$. (Right panel) Inward propagating modes $\omega_-$. The $|\mathrm{Re}[\omega_\pm]|$ (solid lines) give the wave speed $(\omega_\pm /k)$ while $|\mathrm{Im}[\omega_\pm]|$ (dashed lines) sets the damping rate. At small $\varv_A/\varv_0 < 0.01$ (red, solid and red, dashed line) a splitting of inward and outward modes occurs while the phase speed of waves tends to zero. All curves are shown only for the case $k_3 = k$, i.e.~a perturbation along the mean magnetic field $(\theta = 0^\circ)$.}
    \label{fig:bridginglaw}
\end{figure*}

To gain insight into the coupling between the radiation and the magnetic field we here establish a general dependence of the dispersion relation on perturbation wavelength. Under our adapted assumptions, a magneto-radiative coupling only occurs in two eigentensor elements in Eq.~\eqref{eq:eigentensor} (without loss of generality we focus on $\mathcal{M}_{22}$). 

The general dispersion relation \eqref{eq:MHDdispersionrelation} in the radial and lateral direction becomes a quadratic polynomial with eigenmode solutions
\begin{equation}\label{eq:solution_dispersion_radial}
\omega_{33}= 0, \qquad \omega_{33} = i\frac{\delta g_3}{\delta \varv_3}
\end{equation}
\begin{equation}\label{eq:solution_dispersion_hori}
\omega_{22,\pm} = \frac{1}{2} \left[ i\left( \frac{\delta g_2}{\delta \varv_2} \right) \pm \sqrt{4\varv_A^2 k_3^2 - \left( \frac{\delta g_2}{\delta \varv_2}\right)^2} \right],
\end{equation}
where we point out that most of the elements $\delta g_i/\delta \varv_j =0$ if $i\neq j$. The latter follows from symmetry arguments of the radiation in the perturbed radiative acceleration tensor \eqref{eq:DgDvFinal}. Below this will be explicitly demonstrated when we study long- and short-wavelength perturbations. We now discuss the solutions in the radial and horizontal direction separately.

\subsubsection{Radial radiation} 

In the radial direction no coupling between the magnetic field and the radiation occurs (cf.~Eq.~\eqref{eq:eigentensor}). The perturbed radiative acceleration tensor becomes
\begin{equation}\label{eq:bridging_sol_rad}
\frac{\delta g_3}{\delta \varv_3} = \omega_\star \frac{ik_3/\chi_\star}{1+ ik_3/\chi_\star} - \omega_\star \left( \frac{1+\sigma}{1+\sigma/3} \right) \left\langle \frac{i\mu^3 k_3/\chi_\star}{1+ i \mu k_3 /Q_0} \right\rangle,
\end{equation}
where $\sigma$ is related to the wind expansion (see Eq.~\eqref{eq:homowindexp}). The first term on the right-hand side comes from pure absorption as found by \citetalias{1984ApJ...284..337O} while the second term on the right-hand side now also takes into account line-drag. Although this solution is for a point star, we refrain ourselves from further discussing it, because its properties are quite similar to that of a finite-coned star \citepalias{1985ApJ...299..265O}.

\subsubsection{Horizontal radiation}

The perturbed radiative acceleration tensor is
\begin{equation}\label{eq:bridging_sol_hor}
\frac{\delta g_2}{\delta \varv_2} = - \frac{\omega_\star}{2}\left( \frac{1+\sigma}{1+\sigma/3} \right) \left\langle \frac{(1-\mu^2)i\mu k_3/\chi_\star}{1+ i \mu k_3 /Q_0} \right\rangle
\end{equation}
and $\delta g_1/\delta \varv_1$, i.e.~radiation contributions in eigentensor element $\mathcal{M}_{11}$ that describes the fast mode, is exactly the same when making the substitution $k_3 \rightarrow k$, and $k = \sqrt{k_1^2 + k_3^2}$. 

Fig.~\ref{fig:bridginglaw} displays the behaviour of the real and imaginary wave frequency versus wave number for a range of values of the ratio of the Alfv\'en speed $\varv_A$ to the radiative wave speed $\omega_\star/\chi_\star$. The latter is approximately equal to the mean flow (Abbott) speed $\varv_0 \equiv \omega_\star/\chi_\star$. 

%
%

Contrary to a non-magnetic line-driven wind, where the Abbott speed is much higher than the sound speed ($\varv_0 \gg a$, i.e.~a supersonic flow), in the case of a magnetic wind $\varv_A > \varv_0$, $\varv_A \simeq \varv_0$, or $\varv_A < \varv_0$ depending on the local conditions of magnetic field and density. Therefore, in a magnetic line-driven wind any of the curves shown in Fig.~\ref{fig:bridginglaw} is a viable solution to describe the magneto-radiative wave propagation and instability. This is quite different from the non-magnetic radiative-acoustic case studied in \citetalias{1984ApJ...284..337O}.

At very high Alfv\'en speeds $(\varv_A \gg \varv_0)$ curves of $\mathrm{Re}[\omega_\pm]$ are non-dispersive ordinary Alfv\'en waves over the full wavelength range. When $\varv_A/\varv_0$ is lowered ordinary Alfv\'en waves remain if $|\mathrm{Re}[\omega_\pm]| > \omega_\star$ but are modified by the radiation force for $|\mathrm{Re}[\omega_\pm]| < \omega_\star$ and become dispersive magneto-radiative waves instead. In particular, when $\varv_A \sim 0.01\varv_0$ waves with wavelengths near the bridging length $(k\sim \chi_\star)$ are modified by the radiative force such that their propagation characteristics become more complicated. When $\varv_A < 0.01\varv_0$ magneto-radiative waves are strongly modified in their propagation.

An important result of Fig.~\ref{fig:bridginglaw} can be realized by further considering $\mathrm{Im}[\omega_\pm]$. All modes undergo \emph{net damping}, i.e. they have a negative growth rate (this appears positive in the figure because we are plotting $|\mathrm{Im}[\omega_\pm]|$ on a logarithmic scale), which follows directly from the fact that $\delta g_2/\delta \varv_2 <0$ for all perturbation wavelengths in Eq.~\eqref{eq:bridging_sol_hor}. For $\varv_A \gtrsim 0.01\varv_0$ the damping is the same for all modes while it is slightly modified for $\varv_A \lesssim 0.01\varv_0$. Specifically, in the latter case the inward and outward modes become distinct and have different damping properties. The overall effect still remains and all modes undergo net damping. Finally, irrespective of $\varv_A /\varv_0$, the damping rate becomes vanishingly small at long-wavelengths such that the magneto-radiative modes become marginally stable.

It is interesting to note that the ratio $\varv_A/\varv_0$ has to reduce to order 0.01 for a change in mode behaviour to occur. Examining the radical of Eq.~\eqref{eq:solution_dispersion_hori} shows that such situation means that $\varv_A/\varv_0 \sim 1/k$. Since the mode switch manifests itself near $(k\sim \chi_\star)$ this indicates that $\varv_A/\varv_0$ is effectively approaching the bridging length. The latter is a characteristic length of the wind and is on the order of $L_\mathrm{Sob} \approx 0.01R_\star$. This shows that for a majority of expected conditions in the winds of magnetic massive stars the damping of short-wavelength magnetic waves should be very effective.

\begin{figure}
\centering
	\includegraphics[scale=0.7]{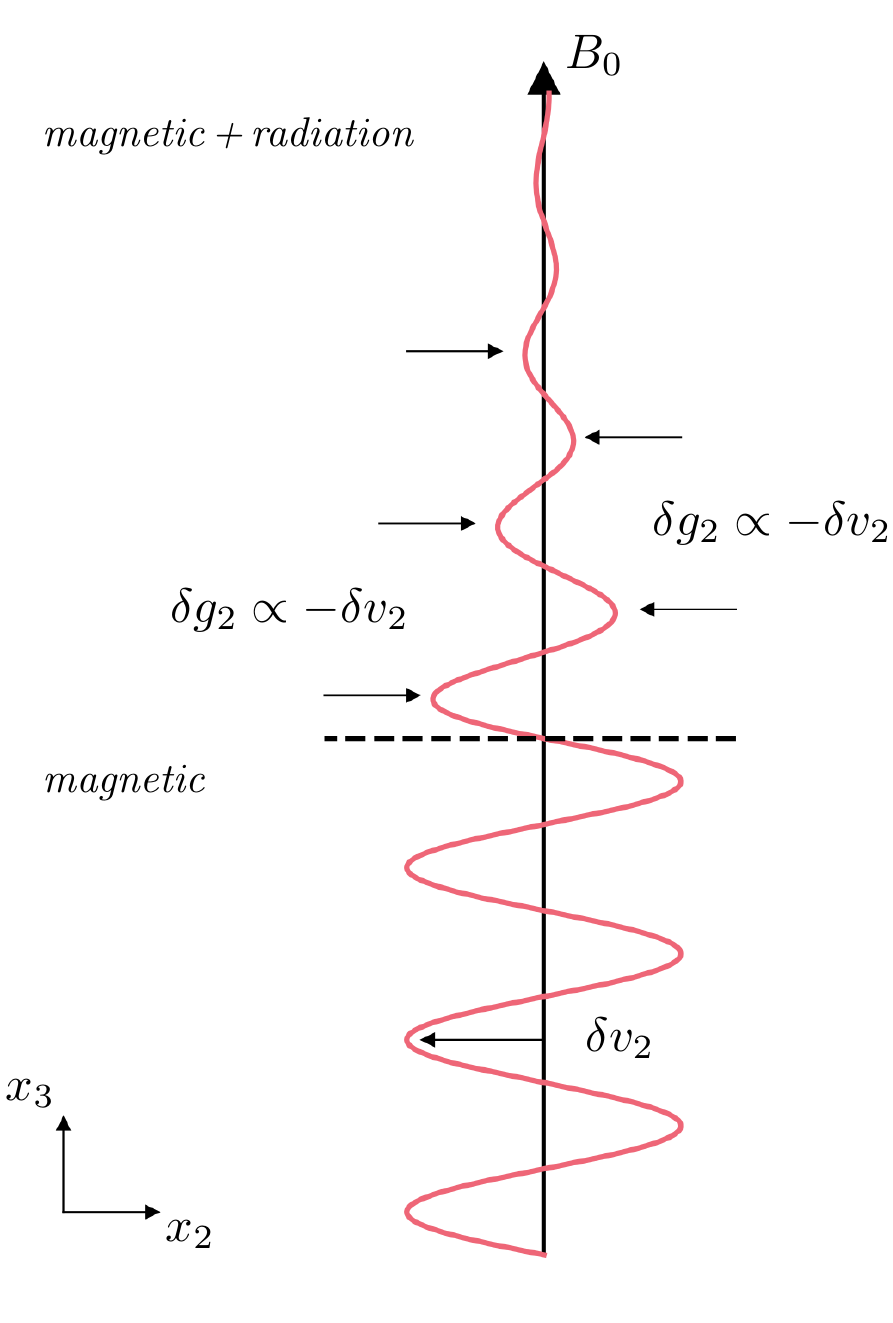}
    \caption{Cartoon depicting the propagation of a short-wavelength Alfv\'en wave resulting from a horizontal velocity perturbation $\delta \varv_2$ along the radial mean magnetic field. (Above dashed line) Damping of the short-wavelength Alfv\'en wave due to the line-drag $\delta g_2/\delta \varv_2$. (Below dashed line) Without radiation field a short-wavelength Alfv\'en wave propagates along the radial magnetic field as a stable transversal wave.}
    \label{fig:damping}
\end{figure}

A key result of this global analysis suggests thus that short-wavelength magnetic waves are always \emph{strongly damped} by radiation. This damping comes from a horizontal line-drag force induced by the radiation onto the magnetic waves as illustrated in Fig.~\ref{fig:damping}. Although the line-drag in reality is much more subtle than this very simple cartoon, the figure nevertheless sketches the basic effect in a somewhat intuitive way. To gain more insight into the underlying physics of the bridging law we now analyse separately the limiting cases of long- and short-wavelength perturbations. 

\subsection{Long-wavelength perturbations}\label{sec:longwav}

\subsubsection{Perturbed radiative acceleration tensor}

The non-magnetic limit of long-wavelength perturbations $(\lambda \gg L_\mathrm{Sob})$ has first been investigated by \citet{1980ApJ...242.1183A}. In the present notation the long-wavelength limit is $k_l \ll Q_0$ making the denominator $(1 + i n_l k_l/Q_0)$ in Eq.~\eqref{eq:DgDvFinal} approach unity. Therefore, the radiative acceleration tensor is
\begin{equation}\label{eq:DgDv_LongWavelength}
\frac{\delta g_i}{\delta \varv_j} = \frac{2\omega_\star}{\chi_\star} \left\langle i n_i n_l k_l n_j \left( \mathcal{D}_\mu -  \frac{\langle \mathcal{D}_\mu p_\mu \rangle}{\langle p_\mu \rangle} \right) \right\rangle.
\end{equation}

The term associated with scattering in the above equation always vanishes in the long-wavelength limit. Specifically, the net angular integrand is odd in the unit normal $n$ such that it vanishes
\begin{equation}
\left\langle n_i n_l k_l n_k \frac{\langle \mathcal{D}_\mu p_\mu \rangle}{\langle p_\mu \rangle} \right\rangle = 0.
\end{equation}

Hence for a pure scattering-driven wind the radiative acceleration tensor becomes
\begin{equation}\label{eq:finalDgDv_ReducedLongWavelength}
\frac{\delta g_i}{\delta \varv_j} = i\frac{2\omega_\star}{\chi_\star} \left\langle  n_i n_l k_l n_j \mathcal{D}_\mu \right\rangle = i \frac{2\omega_\star}{\chi_\star} k_l \mathcal{T}_{ijl}.
\end{equation}

What remains is to solve for the tensor $\mathcal{T}_{ijl}$ which simplifies significantly for a point star. Only radiation directly from the star contributes and this is purely radial and spherically symmetric. Necessarily the two non-radial radiation angles have to vanish, i.e.~$n_1, n_2 \rightarrow 0$. The only remaining component of the radiation field unit normal is the radial direction $n_3$. Recall that $\mathcal{D}_\mu$ is a Dirac delta function such that most terms vanish in the angular integration of $\mathcal{T}_{ijl}$ unless $n_j \parallel r_3$. From this we conclude that $n_i = n_j = n_l = n_3$ and the only non-vanishing tensor component becomes
\begin{equation}
\mathcal{T}_{333} = \langle n_3^3 \delta(\mu -1 ) \rangle = \frac{1}{2}.
\end{equation}

Consequently, the only surviving component of the perturbed radiative acceleration is purely radial
\begin{equation}
\frac{\delta g_3}{\delta \varv_3} = i \frac{\omega_\star}{\chi_\star} k_3.
\end{equation}

This result can also be recovered when considering the general bridging law in the radial direction derived in Eq.~\eqref{eq:bridging_sol_rad}. Indeed, in the long-wavelength limit, $k_3/\chi_\star \ll 1$, the bridging law reduces to the expression derived here.

\subsubsection{Physical interpretation of long-wavelength perturbations}\label{sec:physicslongwav}

The derivation above shows that most radiation associated eigentensor elements in \eqref{eq:eigentensor} vanish such that
\begin{equation}
\begin{aligned}
&\mathcal{M}_{11} = \omega^2 -\varv_A^2 (k_1^2 + k_3^2), \\
&\mathcal{M}_{22} = \omega^2 -\varv_A^2  k_3^2, \\
&\mathcal{M}_{33} = \omega^2 - i \omega \frac{\delta g_3}{\delta \varv_3}, \\
&\mathcal{M}_{ij} = 0, \qquad i\neq j.
\end{aligned}
\end{equation}

The resulting eigenproblem has a unique, non-trivial solution whenever the determinant of the eigentensor is zero
\begin{equation}
\left( \omega^2 + \omega \frac{\omega_\star}{\chi_\star} k \cos \theta  \right) \left( \omega^2 -\varv_A^2 k^2 \right) \left( \omega^2 -\varv_A^2k^2 \cos^2 \theta \right) = 0,
\end{equation}
where we have substituted the expressions $k_1 = k\sin \theta$, $k_3 = k\cos \theta$, and that $k = \sqrt{k_1^2 + k_3^2}$. This sextic polynomial with real coefficients has six independent real solutions:
\begin{equation}
\begin{aligned}
&\omega^{(1)} = -\varv_0 k \cos \theta, \qquad \omega^{(2)} = 0, \qquad \omega^{(3)}_+ = \varv_A k \cos \theta, \\
&\omega^{(4)}_-  = - \varv_A k \cos \theta , \qquad \omega^{(5)}_+ = \varv_A k, \qquad \omega^{(6)}_- = -\varv_A k,
 \end{aligned}
\end{equation}
and the ratio of growth rate to bridging length $\omega_\star /\chi_\star$ yields the mean flow speed $\varv_0$. All the derived eigenfrequencies are real meaning that the corresponding waves are stable. All eigenfrequencies also linearly depend on wave number such that the waves are of non-dispersive type. 

We find one eigenfrequency $-\varv_0 k \cos \theta$ with eigenvector $(0,0,1)$ representing a velocity fluctuation polarised in the parallel direction. This is the Abbott wave propagating inward at the mean flow speed \citep{1980ApJ...242.1183A}. Additionally, four magnetic waves are found with eigenfrequencies $\pm \varv_A k \cos \theta$ and $\pm \varv_A k$ that describe inward/outward propagating Alfv\'en and fast MHD waves with eigenvectors $(0,1,0)$ and $(1,0,0)$, respectively. All these magnetic waves have velocity fluctuations that are horizontally polarised compared to the mean magnetic field. Finally, we find a genuine neutral mode with eigenvector $(0,0,0)$ that involves a density perturbation alone. Interestingly, these results also show that for long-wavelength perturbations there is a complete decoupling between the radiative and magnetic waves such that they do not interact with each other directly. Notice that when we take only radially propagating waves ($k_1 = 0$) then the fast wave vanishes (solutions 5 \& 6) and becomes an Alfv\'en wave. In this case, therefore, a degenerate pair of inward and outward Alfv\'en waves remains together with the inward Abbott wave. On the other hand, in the limit of vanishing magnetic field ($\varv_A\rightarrow0$) only an Abbott wave remains as found by \citet{1980ApJ...242.1183A}.

\subsection{Short-wavelength perturbations}\label{sec:shortwav}

\subsubsection{Perturbed radiative acceleration tensor}

Initial stability considerations of the radiation-hydrodynamic equations of a line-driven wind have been performed by \citet{1979ApJ...231..514M} and \citet{1980ApJ...241.1131C} in the short-wavelength limit ($\lambda \ll L_\mathrm{Sob}$). In terms of the radiative acceleration tensor in this limit the scattering term will not vanish in Eq.~\eqref{eq:DgDvFinal}. Indeed, for short-wavelength perturbations $(k_l \gg Q_0)$ 
\begin{equation}
\frac{\delta g_i}{\delta \varv_j} = \frac{2\omega_\star}{\chi_\star} \left\langle n_i n_j\left( \mathcal{D}_\mu - \frac{\langle \mathcal{D}_\mu p_\mu \rangle}{\langle p_\mu \rangle}  \right) Q_0 \right\rangle = \frac{2\omega_\star}{\chi_\star} \mathcal{T}_{ij}.
\end{equation}

The non-vanishing of scattering contributions at short-wavelengths induces some properties onto the tensor $\mathcal{T}_{ij}$: (i) because $n$ is a unit vector $n_i n_j = \delta_{ij}$ such that $\mathcal{T}_{ij}$ must be a \emph{diagonal tensor}, (ii) $\mathcal{T}_{ij}$ being a diagonal tensor implies that
\begin{equation}
\mathcal{T}_{ii} =  \left\langle \left(  \mathcal{D}_\mu - \frac{\langle \mathcal{D}_\mu p_\mu \rangle}{\langle p_\mu \rangle} \right) Q_0  \right\rangle = 0
\end{equation}
making $\mathcal{T}_{ij}$ a \emph{traceless}, diagonal tensor, and (iii) radiation has symmetry around the radial axis such that the lateral and azimuthal radiation angles are \emph{indistinguishable}, or
\begin{equation}
\mathcal{T}_{11} = \mathcal{T}_{22} = \frac{1}{2}(\mathcal{T}_{ii} - \mathcal{T}_{33}) = -\frac{1}{2}\mathcal{T}_{33},
\end{equation}
where the latter equality comes from the traceless condition (property ii). From these properties it follows that the radial tensor component
\begin{equation}
\mathcal{T}_{33} = 1 - \frac{(1/3 + \sigma/5) }{1+\sigma/3},
\end{equation}
and so
\begin{equation}\label{eq:horizontal_tensorcomp_shortwav}
\mathcal{T}_{11} = \mathcal{T}_{22} = -\frac{1}{2} \left[ 1 - \frac{(1/3 + \sigma/5) }{1+\sigma/3} \right].
\end{equation}

The surviving perturbed radiative acceleration to perturbed velocity tensor elements in the pure scattering limit are then
\begin{equation}
\frac{\delta g_1}{\delta \varv_1} = -\frac{1}{2}\omega_\star \mathcal{T}_{33}, \qquad \frac{\delta g_2}{\delta \varv_2} = -\frac{1}{2}\omega_\star \mathcal{T}_{33}, \qquad \frac{\delta g_3}{\delta \varv_3} = \omega_\star \mathcal{T}_{33}.
\end{equation}

The non-vanishing horizontal perturbed radiative acceleration terms here leads to the coupling with magnetic waves in the eigentensor and the damping phenomenon as shown in Fig.~\ref{fig:bridginglaw}. The physical interpretation of the damping can be attributed to a line-drag \citep{1984ApJ...284..351L} as captured by the $\mathcal{T}_{ii}$ term. This line-drag \emph{reduces} the instability growth in the radial direction, but \emph{damps} it in the horizontal directions.

\subsubsection{Physical interpretation of short-wavelength perturbations}

The eigenproblem is nearly the same as for the long-wavelength limit, except that all diagonal elements of the eigentensor \eqref{eq:eigentensor} will retain the radiative component
\begin{equation}
\begin{aligned}
&\mathcal{M}_{11} = \omega^2 -\varv_A^2 (k_1^2 + k_3^2) - i \omega \frac{\delta g_1}{\delta \varv_1}, \\
&\mathcal{M}_{22} = \omega^2 -\varv_A^2  k_3^2 	- i \omega \frac{\delta g_2}{\delta \varv_2}, \\	
&\mathcal{M}_{33} = \omega^2 - i \omega \frac{\delta g_3}{\delta \varv_3}, \\
&\mathcal{M}_{ij} = 0, \qquad i\neq j.
\end{aligned}
\end{equation}

This coupling between the radiation and the magnetic field has important effects as we will show below. Setting the determinant of the eigentensor to zero to retrieve the non-trivial solutions yields
\begin{equation}\label{eq:determinant_shortwav}
\begin{aligned}
\left( \omega^2 - i\omega_\star \mathcal{T}_{33}\omega \right) &\left( \omega^2 - \varv_A^2 k^2 + i\omega \frac{1}{2}\omega_\star \mathcal{T}_{33} \right) \\
&\times \left(\omega^2 - \varv_A^2 k^2 \cos^2 \theta + i\omega  \frac{1}{2}\omega_\star \mathcal{T}_{33} \right) = 0,
\end{aligned}
\end{equation}
where we have substituted the expressions $k_1 = k\sin \theta$, $k_3 = k\cos \theta$, and that $k = \sqrt{k_1^2 + k_3^2}$. This is a sextic polynomial with mixed real and imaginary coefficients so it has six roots that can be real, imaginary, or appearing in complex conjugate pairs. 

Instead of writing out the eigenmodes of Eq.~\eqref{eq:determinant_shortwav} it is more insightful to make the substitution $w = -i \omega/(\omega_\star \mathcal{T}_{33}) = \mathrm{Im}[\omega]/(\omega_\star \mathcal{T}_{33})$ and $K=\varv_A k/(\omega_\star \mathcal{T}_{33})$ giving
\begin{equation}\label{eq:shortwaveq_dimless}
w(w-1) \left( w^2 +\frac{1}{2}w + K^2 \cos^2 \theta \right) \left( w^2 + \frac{1}{2}w + K^2 \right) = 0.
\end{equation}

\begin{figure*}
	\includegraphics[width=\textwidth]{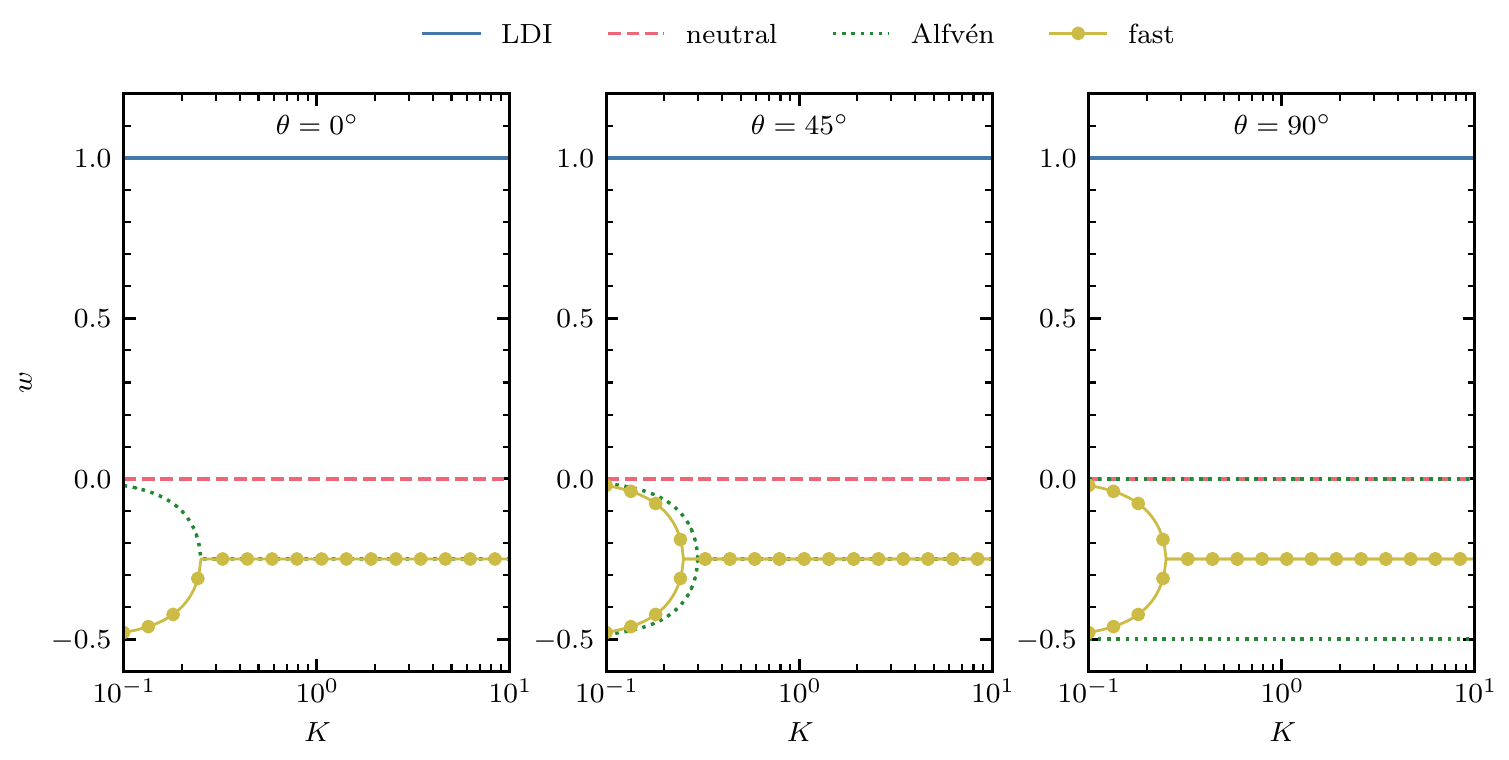}
    \caption{Growth rate $w$ (in units of $\omega_\star \mathcal{T}_{33}$) for magneto-radiative modes as function of wave number $K$ (in units of $\omega_\star \mathcal{T}_{33}/\varv_A$). All panels show a different orientation of the initial wave propagation vector for each mode. At $\theta = 0^\circ$ the Alfv\'en and fast mode overlap before the collapse of the pitchfork branches. For clarity only half of each mode branch is shown here. At $\theta = 90^\circ$ the Alfv\'en mode vanishes leading to an additional neutral mode and a radiative mode.}
    \label{fig:allangles}
\end{figure*}

In Fig.~\ref{fig:allangles} the solution of the sextic equation in $w$ is shown over a range of wave numbers $K$ for each mode. There is always one radial radiative mode (LDI) from a radial velocity polarisation that is \emph{unstable} $(w>0)$ for any angle $\theta$ alongside a neutral, non-propagating density mode. It can be seen that the magnetic modes (Alfv\'en and fast) resulting from a horizontal velocity polarisation experience \emph{net damping} for any angle $\theta$ $(w<0)$. 

The solution curves of the magnetic modes assume a pitchfork shape that is set by a critical perturbation wavelength $K_\mathrm{crit}$. This critical perturbation wavelength essentially sets the behaviour of inward and outward magnetic modes. For $K > K_\mathrm{crit}$ all magnetic modes are damped at the damping rate of the line-drag while inward and outward modes propagate identically. These propagation characteristics are only modified once $K < K_\mathrm{crit}$ and the inward and outward modes become distinct (see also discussion around Fig.~\ref{fig:bridginglaw}). In light of the range of validity of Eq.~\eqref{eq:determinant_shortwav}, i.e.~the short-wavelength limit $(k\gg Q_0)$, and that $Q_0 \sim \chi_\star$ and $\omega_\star \mathcal{T}_{33} \approx \omega_\star$ the wave number variable $K$ satisfies $K\gg \varv_A/\varv_0$. Thus depending on the local conditions in the wind, part of or the complete \emph{magnetic} solution curves in Fig.~\ref{fig:allangles} apply to describe the magnetic mode damping.

The results here thus confirm the bridging law that short-wavelength magnetic waves with horizontal velocity polarisations undergo strong damping due to the line-drag. Overall the wind remains highly unstable, and the most unstable mode has the most radial velocity polarisation. The exact effects of the lateral damping in regulating the wind dynamics and wind fragmentation are speculative and have to be further investigated by means of numerical simulation.

\section{Conclusions \& Future outlook}\label{sec:finals}

In this paper we have studied the linear stability of a line-scattering-driven magnetic stellar wind. We have considered both radial and non-radial wave propagation in the limiting cases of long-wavelength and short-wavelength compared to the Sobolev length of the wind. In addition, we have also investigated the dependency on perturbation wavelength by establishing a bridging law. Within our assumption of a radial magnetic field we have found that stable, long-wavelength radiative and magnetic waves are decoupled. A key result is that short-wavelength Alfv\'en waves are strongly damped due to a drag force from line-scattering. Additionally, when the ratio of Alfv\'en speed to radiative wave speed becomes small and approaches a scale comparable to the Sobolev length, the damping properties of magnetic waves are slightly modified. This situation is more likely to play a role in weakly magnetic environments, or in a very dense wind where low Alfv\'en speeds can occur. Nevertheless, the overall result remains that short-wavelength magnetic waves always undergo damping.

Possible analytical extensions to the analysis here include introducing thermal pressure or considering non-radial magnetic fields. Since the mean flow speed $\varv_0 \gg a$ in most parts of line-driven winds, such sound speed terms are typically not important for the dynamics of the flow. However, in a magnetic line-driven wind the stability properties may also depend on the comparison between the Alfv\'en speed and the sound speed. This could potentially alter the instability growth and damping for short-wavelength perturbations. The effect of non-radial mean magnetic field components might potentially alter the coupling properties of radiative and magnetic waves and their linear (in)stability. Such non-radial magnetic field components can occur in rapidly rotating magnetic massive stars or in regions of strong field line curvature in the circumstellar magnetosphere near the magnetic equatorial plane.

Furthermore, we plan to present non-linear magnetohydrodynamic instability simulations that study the development and dynamics of unstable LDI structures. These simulations aim to investigate both the \emph{global} and \emph{local} dynamics. In particular, the latter will be able to study the damping of Alfv\'en waves as found in the present paper with the inclusion of the lateral perturbed radiative acceleration. Such physics might prevent, or alter, the lateral fragmentation seen in non-magnetic 2D instability simulations \citep{2018A&A...611A..17S} and affect the wind clumping properties. The inclusion of magnetic fields might also affect the growth rate of the instability and the wave-stretching mechanism \citep{2017MNRAS.469.3102F} that causes the separation between clumps and the interclump medium. Our first tentative global simulations already suggest that lateral break-up of wind material is strongly modified depending on the background stellar magnetic field strength (Driessen et.~al., in prep.). The instability dynamics in such a setting and its effect on observational diagnostics will be reported in a future work.

\section*{Acknowledgements}

FAD thanks Jonas Berx for sharing some of his Mathematica tricks useful for this work. FAD and JOS acknowledge support from the Odysseus program of the Belgian Research Foundation Flanders (FWO) under grant G0H9218N. NDK acknowledges support from the KU Leuven C1 grant MAESTRO C16/17/007. We thank the referee, Prof.~Achim Feldmeier, for constructive criticism and comments to the manuscript.

\section*{Data availability}

The data underlying this article are available in the article.




\bibliographystyle{mnras}
\bibliography{refs} 

\begin{thebibliography}{}
\makeatletter
\relax
\def\mn@urlcharsother{\let\do\@makeother \do\$\do\&\do\#\do\^\do\_\do\%\do\~}
\def\mn@doi{\begingroup\mn@urlcharsother \@ifnextchar [ {\mn@doi@}
  {\mn@doi@[]}}
\def\mn@doi@[#1]#2{\def\@tempa{#1}\ifx\@tempa\@empty \href
  {http://dx.doi.org/#2} {doi:#2}\else \href {http://dx.doi.org/#2} {#1}\fi
  \endgroup}
\def\mn@eprint#1#2{\mn@eprint@#1:#2::\@nil}
\def\mn@eprint@arXiv#1{\href {http://arxiv.org/abs/#1} {{\tt arXiv:#1}}}
\def\mn@eprint@dblp#1{\href {http://dblp.uni-trier.de/rec/bibtex/#1.xml}
  {dblp:#1}}
\def\mn@eprint@#1:#2:#3:#4\@nil{\def\@tempa {#1}\def\@tempb {#2}\def\@tempc
  {#3}\ifx \@tempc \@empty \let \@tempc \@tempb \let \@tempb \@tempa \fi \ifx
  \@tempb \@empty \def\@tempb {arXiv}\fi \@ifundefined
  {mn@eprint@\@tempb}{\@tempb:\@tempc}{\expandafter \expandafter \csname
  mn@eprint@\@tempb\endcsname \expandafter{\@tempc}}}

\bibitem[\protect\citeauthoryear{{Abbott}}{{Abbott}}{1980}]{1980ApJ...242.1183A}
{Abbott} D.~C.,  1980, \mn@doi [\apj] {10.1086/158550}, \href
  {http://adsabs.harvard.edu/abs/1980ApJ...242.1183A} {242, 1183}

\bibitem[\protect\citeauthoryear{{Berghoefer}, {Schmitt}, {Danner}  \&
  {Cassinelli}}{{Berghoefer} et~al.}{1997}]{1997A&A...322..167B}
{Berghoefer} T.~W.,  {Schmitt} J.~H.~M.~M.,  {Danner} R.,   {Cassinelli} J.~P.,
   1997, \aap, \href {http://adsabs.harvard.edu/abs/1997A%26A...322..167B}
  {322, 167}

\bibitem[\protect\citeauthoryear{{Carlberg}}{{Carlberg}}{1980}]{1980ApJ...241.1131C}
{Carlberg} R.~G.,  1980, \mn@doi [\apj] {10.1086/158428}, \href
  {http://adsabs.harvard.edu/abs/1980ApJ...241.1131C} {241, 1131}

\bibitem[\protect\citeauthoryear{{Castor}}{{Castor}}{1970}]{1970MNRAS.149..111C}
{Castor} J.~I.,  1970, \mn@doi [\mnras] {10.1093/mnras/149.2.111}, \href
  {https://ui.adsabs.harvard.edu/abs/1970MNRAS.149..111C} {149, 111}

\bibitem[\protect\citeauthoryear{{Castor}, {Abbott}  \& {Klein}}{{Castor}
  et~al.}{1975}]{1975ApJ...195..157C}
{Castor} J.~I.,  {Abbott} D.~C.,   {Klein} R.~I.,  1975, \mn@doi [\apj]
  {10.1086/153315}, \href {http://adsabs.harvard.edu/abs/1975ApJ...195..157C}
  {195, 157}

\bibitem[\protect\citeauthoryear{{Dessart} \& {Owocki}}{{Dessart} \&
  {Owocki}}{2002}]{2002A&A...383.1113D}
{Dessart} L.,  {Owocki} S.~P.,  2002, \mn@doi [\aap]
  {10.1051/0004-6361:20011826}, \href
  {https://ui.adsabs.harvard.edu/abs/2002A&A...383.1113D} {383, 1113}

\bibitem[\protect\citeauthoryear{{Donati}, {Babel}, {Harries}, {Howarth},
  {Petit}  \& {Semel}}{{Donati} et~al.}{2002}]{2002MNRAS.333...55D}
{Donati} J.-F.,  {Babel} J.,  {Harries} T.~J.,  {Howarth} I.~D.,  {Petit} P.,
  {Semel} M.,  2002, \mn@doi [\mnras] {10.1046/j.1365-8711.2002.05379.x}, \href
  {http://adsabs.harvard.edu/abs/2002MNRAS.333...55D} {333, 55}

\bibitem[\protect\citeauthoryear{{Driessen}, {Sundqvist}  \& {Kee}}{{Driessen}
  et~al.}{2019}]{2019A&A...631A.172D}
{Driessen} F.~A.,  {Sundqvist} J.~O.,   {Kee} N.~D.,  2019, \mn@doi [\aap]
  {10.1051/0004-6361/201936331}, \href
  {https://ui.adsabs.harvard.edu/abs/2019A&A...631A.172D} {631, A172}

\bibitem[\protect\citeauthoryear{{Eversberg}, {L{\'e}pine}  \&
  {Moffat}}{{Eversberg} et~al.}{1998}]{1998ApJ...494..799E}
{Eversberg} T.,  {L{\'e}pine} S.,   {Moffat} A.~F.~J.,  1998, \mn@doi [\apj]
  {10.1086/305218}, \href {http://adsabs.harvard.edu/abs/1998ApJ...494..799E}
  {494, 799}

\bibitem[\protect\citeauthoryear{{Feldmeier} \& {Thomas}}{{Feldmeier} \&
  {Thomas}}{2017}]{2017MNRAS.469.3102F}
{Feldmeier} A.,  {Thomas} T.,  2017, \mn@doi [\mnras] {10.1093/mnras/stx858},
  \href {https://ui.adsabs.harvard.edu/abs/2017MNRAS.469.3102F} {469, 3102}

\bibitem[\protect\citeauthoryear{{Feldmeier}, {Puls}  \&
  {Pauldrach}}{{Feldmeier} et~al.}{1997}]{1997A&A...322..878F}
{Feldmeier} A.,  {Puls} J.,   {Pauldrach} A.~W.~A.,  1997, \aap, \href
  {https://ui.adsabs.harvard.edu/abs/1997A&A...322..878F} {322, 878}

\bibitem[\protect\citeauthoryear{{Fossati} et~al.,}{{Fossati}
  et~al.}{2015}]{2015A&A...582A..45F}
{Fossati} L.,  et~al., 2015, \mn@doi [\aap] {10.1051/0004-6361/201526725},
  \href {https://ui.adsabs.harvard.edu/abs/2015A&A...582A..45F} {582, A45}

\bibitem[\protect\citeauthoryear{{Lucy}}{{Lucy}}{1983}]{1983ApJ...274..372L}
{Lucy} L.~B.,  1983, \mn@doi [\apj] {10.1086/161453}, \href
  {http://adsabs.harvard.edu/abs/1983ApJ...274..372L} {274, 372}

\bibitem[\protect\citeauthoryear{{Lucy}}{{Lucy}}{1984}]{1984ApJ...284..351L}
{Lucy} L.~B.,  1984, \mn@doi [\apj] {10.1086/162413}, \href
  {https://ui.adsabs.harvard.edu/abs/1984ApJ...284..351L} {284, 351}

\bibitem[\protect\citeauthoryear{{Lucy} \& {Solomon}}{{Lucy} \&
  {Solomon}}{1970}]{1970ApJ...159..879L}
{Lucy} L.~B.,  {Solomon} P.~M.,  1970, \mn@doi [\apj] {10.1086/150365}, \href
  {https://ui.adsabs.harvard.edu/abs/1970ApJ...159..879L} {159, 879}

\bibitem[\protect\citeauthoryear{{MacGregor}, {Hartmann}  \&
  {Raymond}}{{MacGregor} et~al.}{1979}]{1979ApJ...231..514M}
{MacGregor} K.~B.,  {Hartmann} L.,   {Raymond} J.~C.,  1979, \mn@doi [\apj]
  {10.1086/157213}, \href {http://adsabs.harvard.edu/abs/1979ApJ...231..514M}
  {231, 514}

\bibitem[\protect\citeauthoryear{{Owocki} \& {Puls}}{{Owocki} \&
  {Puls}}{1996}]{1996ApJ...462..894O}
{Owocki} S.~P.,  {Puls} J.,  1996, \mn@doi [\apj] {10.1086/177203}, \href
  {http://adsabs.harvard.edu/abs/1996ApJ...462..894O} {462, 894}

\bibitem[\protect\citeauthoryear{{Owocki} \& {Puls}}{{Owocki} \&
  {Puls}}{1999}]{1999ApJ...510..355O}
{Owocki} S.~P.,  {Puls} J.,  1999, \mn@doi [\apj] {10.1086/306561}, \href
  {https://ui.adsabs.harvard.edu/abs/1999ApJ...510..355O} {510, 355}

\bibitem[\protect\citeauthoryear{{Owocki} \& {Rybicki}}{{Owocki} \&
  {Rybicki}}{1984}]{1984ApJ...284..337O}
{Owocki} S.~P.,  {Rybicki} G.~B.,  1984, \mn@doi [\apj] {10.1086/162412}, \href
  {http://adsabs.harvard.edu/abs/1984ApJ...284..337O} {284, 337}

\bibitem[\protect\citeauthoryear{{Owocki} \& {Rybicki}}{{Owocki} \&
  {Rybicki}}{1985}]{1985ApJ...299..265O}
{Owocki} S.~P.,  {Rybicki} G.~B.,  1985, \mn@doi [\apj] {10.1086/163697}, \href
  {http://adsabs.harvard.edu/abs/1985ApJ...299..265O} {299, 265}

\bibitem[\protect\citeauthoryear{{Owocki} \& {Rybicki}}{{Owocki} \&
  {Rybicki}}{1986}]{1986ApJ...309..127O}
{Owocki} S.~P.,  {Rybicki} G.~B.,  1986, \mn@doi [\apj] {10.1086/164586}, \href
  {https://ui.adsabs.harvard.edu/abs/1986ApJ...309..127O} {309, 127}

\bibitem[\protect\citeauthoryear{{Owocki} \& {Rybicki}}{{Owocki} \&
  {Rybicki}}{1991}]{1991ApJ...368..261O}
{Owocki} S.~P.,  {Rybicki} G.~B.,  1991, \mn@doi [\apj] {10.1086/169689}, \href
  {https://ui.adsabs.harvard.edu/abs/1991ApJ...368..261O} {368, 261}

\bibitem[\protect\citeauthoryear{{Owocki}, {Castor}  \& {Rybicki}}{{Owocki}
  et~al.}{1988}]{1988ApJ...335..914O}
{Owocki} S.~P.,  {Castor} J.~I.,   {Rybicki} G.~B.,  1988, \mn@doi [\apj]
  {10.1086/166977}, \href
  {https://ui.adsabs.harvard.edu/abs/1988ApJ...335..914O} {335, 914}

\bibitem[\protect\citeauthoryear{{Puls}, {Sundqvist}  \& {Markova}}{{Puls}
  et~al.}{2015}]{2015IAUS..307...25P}
{Puls} J.,  {Sundqvist} J.~O.,   {Markova} N.,  2015, in {Meynet} G.,  {Georgy}
  C.,  {Groh} J.,   {Stee} P.,  eds,  IAU Symposium Vol. 307, New Windows on
  Massive Stars. pp 25--36 (\mn@eprint {arXiv} {1409.3582}),
  \mn@doi{10.1017/S174392131400622X}

\bibitem[\protect\citeauthoryear{{Rybicki}, {Owocki}  \& {Castor}}{{Rybicki}
  et~al.}{1990}]{1990ApJ...349..274R}
{Rybicki} G.~B.,  {Owocki} S.~P.,   {Castor} J.~I.,  1990, \mn@doi [\apj]
  {10.1086/168312}, \href
  {https://ui.adsabs.harvard.edu/abs/1990ApJ...349..274R} {349, 274}

\bibitem[\protect\citeauthoryear{{Schneider}, {Ohlmann}, {Podsiadlowski},
  {R{\"o}pke}, {Balbus}, {Pakmor}  \& {Springel}}{{Schneider}
  et~al.}{2019}]{2019Natur.574..211S}
{Schneider} F. R.~N.,  {Ohlmann} S.~T.,  {Podsiadlowski} P.,  {R{\"o}pke}
  F.~K.,  {Balbus} S.~A.,  {Pakmor} R.,   {Springel} V.,  2019, \mn@doi [\nat]
  {10.1038/s41586-019-1621-5}, \href
  {https://ui.adsabs.harvard.edu/abs/2019Natur.574..211S} {574, 211}

\bibitem[\protect\citeauthoryear{{Sobolev}}{{Sobolev}}{1960}]{1960mes..book.....S}
{Sobolev} V.~V.,  1960, {Moving envelopes of stars}.
{Harvard University Press}

\bibitem[\protect\citeauthoryear{{Sundqvist}, {Owocki}  \& {Puls}}{{Sundqvist}
  et~al.}{2012}]{2012ASPC..465..119S}
{Sundqvist} J.~O.,  {Owocki} S.~P.,   {Puls} J.,  2012, in {Drissen} L.,
  {Robert} C.,  {St-Louis} N.,   {Moffat} A.~F.~J.,  eds,  Astronomical Society
  of the Pacific Conference Series Vol. 465, Proceedings of a Scientific
  Meeting in Honor of Anthony F. J. Moffat. p.~119 (\mn@eprint {arXiv}
  {1110.0485})

\bibitem[\protect\citeauthoryear{{Sundqvist}, {Owocki}  \& {Puls}}{{Sundqvist}
  et~al.}{2018}]{2018A&A...611A..17S}
{Sundqvist} J.~O.,  {Owocki} S.~P.,   {Puls} J.,  2018, \mn@doi [\aap]
  {10.1051/0004-6361/201731718}, \href
  {https://ui.adsabs.harvard.edu/abs/2018A&A...611A..17S} {611, A17}

\bibitem[\protect\citeauthoryear{{Wade} et~al.,}{{Wade}
  et~al.}{2016}]{2016MNRAS.456....2W}
{Wade} G.~A.,  et~al., 2016, \mn@doi [\mnras] {10.1093/mnras/stv2568}, \href
  {https://ui.adsabs.harvard.edu/abs/2016MNRAS.456....2W} {456, 2}

\bibitem[\protect\citeauthoryear{{ud-Doula} \& {Owocki}}{{ud-Doula} \&
  {Owocki}}{2002}]{2002ApJ...576..413U}
{ud-Doula} A.,  {Owocki} S.~P.,  2002, \mn@doi [\apj] {10.1086/341543}, \href
  {http://adsabs.harvard.edu/abs/2002ApJ...576..413U} {576, 413}

\bibitem[\protect\citeauthoryear{{ud-Doula}, {Townsend}  \&
  {Owocki}}{{ud-Doula} et~al.}{2006}]{2006ApJ...640L.191U}
{ud-Doula} A.,  {Townsend} R.~H.~D.,   {Owocki} S.~P.,  2006, \mn@doi [\apjl]
  {10.1086/503382}, \href {http://adsabs.harvard.edu/abs/2006ApJ...640L.191U}
  {640, L191}

\bibitem[\protect\citeauthoryear{{ud-Doula}, {Owocki}  \&
  {Townsend}}{{ud-Doula} et~al.}{2008}]{2008MNRAS.385...97U}
{ud-Doula} A.,  {Owocki} S.~P.,   {Townsend} R.~H.~D.,  2008, \mn@doi [\mnras]
  {10.1111/j.1365-2966.2008.12840.x}, \href
  {http://adsabs.harvard.edu/abs/2008MNRAS.385...97U} {385, 97}

\makeatother
\end{thebibliography}



\appendix 

\section{The perturbed radiative acceleration tensor} \label{sec:appendix}


In this Section we perform a detailed derivation on how to obtain the perturbed radiative acceleration tensor in Eq.~\eqref{eq:DgDvFinal}.

\subsection{Equation of radiative transfer}

We describe the radiation field by the spherically symmetric time-independent equation of transfer for a single line in the comoving frame into direction $n_j$ \citepalias{1985ApJ...299..265O} :
\begin{equation}\label{eq:Transferequation}
n_j \frac{\partial I}{\partial r_j}  - Q \frac{\partial I}{\partial x} = -k_L^{(x)}(I - S_L),
\end{equation}
where we neglect aberration and other terms on the order $\mathcal{O}(\varv/c)$. The transfer equation describes the evolution of intensity $I(n_j,r_i,x)$ for a given line-source function $S_L(r_i)$ (we will drop these dependencies again for notational clarity). Radiation propagation is described by a unit vector $n$ with components
\begin{equation}
n_1 = \sin \Theta \cos \Phi, \qquad n_2 = \sin \Theta \sin \Phi, \qquad n_3 = \cos \Theta,
\end{equation}
with $\Theta$ the lateral angle and $\Phi$ the azimuthal angle. We stress that $\Theta$, $\Phi$ are part of the momentum coordinates of the radiation field and unrelated to the real space coordinates $\theta$, $\phi$ of any global background spherical flow.

The normalised comoving frame frequency is
\begin{equation}
x \equiv x_\mathrm{CMF} =  \frac{\nu_\mathrm{CMF} -\nu_0}{\Delta \nu_D}, \qquad \Delta \nu_D = \frac{\nu_0 \varv_\mathrm{th}}{c},
\end{equation}
with $\nu_0$ the line-centre frequency, $\varv_\mathrm{th}$ a thermal speed, $c$ the speed of light, and $\Delta \nu_D$ the Doppler width of a line. 

Line-extinction in the wind is defined according to
\begin{equation}
k_L^{(x)} = k_L \phi(x),
\end{equation}
for a normalised line-profile function
\begin{equation}
 \int_{-\infty}^{+\infty} dx \phi(x) = 1.
\end{equation}

The frequency-integrated line-extinction $k_L$ (in cm$^{-1}$) depends on the upper $u$ and lower level $l$ populations and oscillator strength $f_{lu}$ of the line
\begin{equation}
k_L = \frac{1}{\Delta \nu_D} \sigma_\mathrm{cl} n_l f_{lu},
\end{equation}
for classical cross-section $\sigma_\mathrm{cl}$, number density of the lower level $n_l$, and where we have neglected stimulated emission. Furthermore, $k_L$ can be related to a mass-absorption coefficient $\kappa$ such that $k_L = \kappa \rho$. 

In complete frequency redistribution it follows that for pure scattering
\begin{equation}
S_L = \bar{J} = \langle \bar{I} \rangle,
\end{equation}
where the mean intensity is defined as
\begin{equation}
\bar{J}(r_i) \equiv \frac{1}{2} \int_{-1}^{+1} d\mu \bar{I}(n_j,r_i,x) = \langle \bar{I} \rangle,
\end{equation}
with $\mu = \cos \Theta$, and the frequency-integrated intensity
\begin{equation}
\bar{I}(n_j,r_i) \equiv \int_{-\infty}^{+\infty} dx I(n_j,r_i,x).
\end{equation}

The factor $Q$ in Eq.~\eqref{eq:Transferequation} above contains the projection of the rate of strain tensor $\partial \varv_i /\partial r_j$ into a direction $n_j$
\begin{equation}
Q = \frac{1}{\varv_\mathrm{th}} \frac{\partial \varv_i}{\partial r_j} n_i n_j,
\end{equation}
here it is implicitly assumed that the mean flow is monotonically increasing, i.e.~$Q > 0$. In a spherically symmetric flow the mean state $Q_0$ is
\begin{equation}
Q_0 =  \frac{1}{\varv_\mathrm{th}} \left[ \mu^2\frac{d\varv_0}{dr} + (1-\mu^2)\frac{\varv_0}{r} \right] = \frac{\varv_0}{\varv_\mathrm{th} r}(1+\sigma \mu^2),
\end{equation}

In the latter equality we have introduced the homologeous wind expansion quantity $\sigma \equiv d(\ln \varv_0)/d(\ln r) - 1$ \citep{1970MNRAS.149..111C}. If we describe the mean wind flow by a typical velocity law
\begin{equation}
\varv_0(r) = \varv_\infty \left( 1 - \frac{R_\star}{r} \right)^\beta,
\end{equation}
with $\beta = 0.5$ for a point star and $\varv_\infty$ the terminal wind speed, then it follows that
\begin{equation}\label{eq:homowindexp}
\sigma(r) = \frac{(\beta + 1)R_\star - r}{r-R_\star}.
\end{equation}

The perturbed $\delta Q$, on the other hand, is a truly multidimensional quantity because perturbations are allowed to happen into any direction
\begin{equation}\label{eq:DeltaQ}
\delta Q =  \frac{1}{\varv_\mathrm{th}}  \frac{\partial \delta \varv_i}{\partial r_j} n_i n_j.
\end{equation}

Finally, with these definitions of the radiation field the radiative line-force (per unit mass) becomes
\begin{equation}\label{eq:lineforce}
g_i(r_j) \equiv \frac{4\pi k_L}{\rho c} \bar{H}_i(r_j),
\end{equation}
where $\bar{H}$ is the Eddington line-flux defined as the first moment of the angle-averaged, frequency-integrated intensity
\begin{equation}\label{eq:Eddingtonflux}
\bar{H}_i(r_j) \equiv \langle n_i \bar{I} \rangle.
\end{equation}

We are now ready to determine how linear velocity perturbations affect the radiation field by finding the tensor $\delta g_i/\delta \varv_j$.

\subsection{Linear perturbations in three dimensions}

Following the same procedure as for the magnetic flow, we apply perturbations on Eq.~\eqref{eq:Transferequation} of the form
\begin{equation}
\begin{aligned}
&I \equiv I_0(n_j,r_i,x) + \delta I(n_j,r_i,x), \qquad k_L^{(x)} \equiv k_{L,0}^{(x)}, \\
&S_L \equiv S_{L,0}(r_i) + \delta S_L(r_i), \qquad Q \equiv Q_0(\mu,r) + \delta Q(n_j,r_i).
\end{aligned}
\end{equation}

Up to first order the mean radiation field is described by
\begin{equation}\label{eq:Transfereq_meanstart}
Q_0 \frac{\partial I_0}{\partial x} = k_{L,0}^{(x)} (I_0-S_{L,0}),
\end{equation}
while in the WKB limit and up to first order the perturbed radiation field is
\begin{equation}\label{eq:Transfereq_perturbedstart}
n_j \frac{\partial \delta I}{\partial r_j} - Q_0 \frac{\partial \delta I}{\partial x}  - Q_0 \frac{\partial I_0}{\partial x} = -k_{L,0}^{(x)} (\delta I-  \delta S_L).
\end{equation}

\subsection{Solution of the mean radiation field}

To solve for the mean radiation field in Eq.~\eqref{eq:Transfereq_meanstart} a suitable (blue-wing) boundary condition is
\begin{equation}
I_0(n_j,r_i,x\rightarrow +\infty) = I_\star \mathcal{D}(n_j,r_i),
\end{equation}
where $I_\star$ is the stellar core intensity and $\mathcal{D}(n_j,r_i) \equiv \mathcal{D}_\mu$ is an angular function capturing possible intensity variations across the stellar disk (a Dirac delta function for a point star). As the Sobolev approximation holds for the mean flow the solution of Eq.~\eqref{eq:Transfereq_meanstart} is simplified when introducing the angle-dependent Sobolev optical depth $\tau_\mu = k_L/Q_0$ or
\begin{equation}
\tau_\mu = \frac{\tau_{\mu=0} }{1+\sigma \mu^2},
\end{equation}
where $\tau_{\mu=0} = k_L \varv_\mathrm{th} r /\varv_0$ is the Sobolev optical depth in the lateral direction. Furthermore, we define the integration function
\begin{equation}
\Phi(x) \equiv \int_x^{+\infty} dx' \phi(x')
\end{equation}
so that 
\begin{equation}
\tau_\mu \Phi(x) = \frac{k_L}{Q_0} \int_x^{+\infty} dx' \phi(x') = \frac{1}{Q_0} \int_x^{+\infty} dx' k_{L,0}^{(x')}.
\end{equation}

The mean radiation field, Eq.~\eqref{eq:Transfereq_meanstart}, can be solved when using an integrating factor $\exp[-\tau_\mu\Phi(x)]$
\begin{equation}\label{eq:Transfersolution_meanstate}
I_0 = I_\star  \mathcal{D}_\mu e^{-\tau_\mu \Phi(x)} + S_{L,0} (1-e^{-\tau_\mu \Phi(x)}).
\end{equation}

We proceed to determine the mean Eddington line-flux by computing the frequency-integrated mean intensity and insert this condition into Eq.~\eqref{eq:Transfersolution_meanstate}. It follows that
\begin{equation}
\bar{I}_0 = I_\star \mathcal{D}_\mu p_\mu + S_{L,0}(1 - p_\mu),
\end{equation}
where we define the escape probability of a single line as
\begin{equation}
p_\mu \equiv \int_{-\infty}^{+\infty} dx \phi(x) e^{-\tau_\mu\Phi(x)} = \frac{1-e^{-\tau_\mu}}{\tau_\mu}.
\end{equation}

Taking an angular average of the frequency-integrated mean intensity yields
\begin{equation}\label{eq:Meanradfield_angle+freqaverage}
\bar{J} = \langle \bar{I}_0 \rangle = I_\star \langle \mathcal{D}_\mu p_\mu \rangle + S_{L,0} \left( 1 - \langle p_\mu \rangle \right),
\end{equation}
and by virtue of Eq.~\eqref{eq:Eddingtonflux} it follows that
\begin{equation}
\bar{H}_{i0} =I_\star \langle n_i \mathcal{D}_\mu p_\mu \rangle.
\end{equation}

Notice that in the latter expression any dependence on the angle-dependent line-source function has vanished. This follows from the fact that in the Sobolev approximation the escape probability (and so the Sobolev optical depth) is an even function in the vector $n$. Therefore, in the mean flow the diffuse flux vanishes and the Eddington line-flux is solely due to absorption of the direct component of intensity.

Finally, the mean radiative line-force follows from Eq.~\eqref{eq:lineforce}
\begin{equation}
g_0 = \left( \frac{4\pi k_L}{\rho_0 c} \right) \bar{H}_0 = \left( \frac{4\pi \kappa I_\star}{c} \right) \langle \mu \mathcal{D}_\mu p_\mu \rangle,
\end{equation}
where we have used the assumption of radially streaming radiation.

\subsection{Solution of the perturbed radiation field}

For the solution of the perturbed transfer equation, Eq.~\eqref{eq:Transfereq_perturbedstart}, a complication occurs due to the appearance of $\partial I_0/\partial x$ which itself depends on the mean state line-source function $S_0$ (see Eq.~\eqref{eq:Transfereq_meanstart}). Indeed, $\partial I_0/\partial x \propto (I_0 - S_{L,0})$ which makes the perturbed transfer equation \eqref{eq:Transfereq_perturbedstart}
\begin{equation}
\frac{1}{Q_0} n_j \frac{\partial \delta I}{\partial r_j} - \frac{\partial \delta I}{\partial x} + \tau_\mu\delta I = \tau_\mu\left( \delta S_L + (I_0 - S_{L,0}) \frac{\delta Q}{Q_0}\right).
\end{equation}

A solution for $I_0$ has been found in Eq.~\eqref{eq:Transfersolution_meanstate}. However, rewriting Eq.~\eqref{eq:Transfersolution_meanstate} shows that $(I_0 - S_{L,0}) \propto -S_{L,0}\exp[ -\tau_\mu \Phi(x) ]$. Therefore, we cannot solve the perturbed transfer equation until an expression for the mean state line-source function $S_{L,0}$ is found. Using again the Sobolev approximation and applying the isotropic scattering condition, $S_{L,0} = \bar{J}$, on Eq.~\eqref{eq:Meanradfield_angle+freqaverage} results in
\begin{equation}\label{eq:line_source_func_expression}
S_{L,0} =  I_\star  \frac{\langle \mathcal{D}_\mu p_\mu \rangle}{\langle p_\mu \rangle},
\end{equation}
which allows us to rewrite Eq.~\eqref{eq:Transfersolution_meanstate} without any additional line-source function dependency 
\begin{equation}
I_0 - S_{L,0} = I_\star e^{-\tau_\mu \Phi(x)} \left( \mathcal{D}_\mu - \frac{\langle \mathcal{D}_\mu p_\mu \rangle}{\langle p_\mu \rangle}  \right).
\end{equation}

Having this expression, the solution of the perturbed radiation field can be found. In analogy with the treatment of the magnetic flow, we apply a plane wave decomposition onto the radiation field, i.e.~$\delta I = \delta \tilde{I} \exp{[i(k_j r_j - \omega t)]}$. The latter will only affect the first term on the left-hand side of Eq.~\eqref{eq:Transfereq_perturbedstart} making a term $i k_j$ appear. The solution strategy is similar to that of the mean radiation field solution with boundary condition,
\begin{equation}
\delta I(n_j,r_i,x\rightarrow +\infty) = 0
\end{equation}
and using an integrating factor $\exp[i n_j k_j x/Q_0 - \tau_\mu\Phi(x)]$ such that
\begin{equation}\label{eq:Transfersolution_perturbstate}
\begin{aligned}
\delta I = {}  &\tau_\mu \delta S_L e^{i(n_j k_j/Q_0)x -\tau_\mu\Phi(x)}  \int_x^{+\infty} dx' \phi(x') e^{-i(n_j k_j/Q_0)x'+ \tau_\mu\Phi(x')} \\
& +  \tau_\mu I_\star \left( \mathcal{D}_\mu -  \frac{\langle \mathcal{D}_\mu p_\mu \rangle}{\langle p_\mu \rangle}  \right) \frac{\delta Q}{Q_0} e^{i(n_j k_j/Q_0)x - \tau_\mu\Phi(x)} \\
&\times \int_x^{+\infty} dx' \phi(x') e^{-i(n_j k_j/Q_0)x'} .
\end{aligned}
\end{equation}

In a similar fashion as in the previous Section we take the frequency-integration of the perturbed intensity
\begin{equation}\label{eq:Perturbradfield_freqaverage}
\delta \bar{I} = \delta S_L \xi  + I_\star \left( \mathcal{D}_\mu -  \frac{\langle \mathcal{D}_\mu p_\mu \rangle}{\langle p_\mu \rangle}  \right) \frac{\delta Q}{Q_0} \zeta,
\end{equation}
with
\begin{equation}
\begin{aligned}
\xi \equiv \tau_\mu  &\int_{-\infty}^{+\infty} dx \phi(x) e^{i(n_j k_j/Q_0)x - \tau_\mu\Phi(x)} \\
&\times \int_x^{+\infty} dx' \phi(x') e^{-i(n_j k_j/Q_0)x' + \tau_\mu\Phi(x')},
\end{aligned}
\end{equation}
\begin{equation}
\begin{aligned}
\zeta \equiv \tau_\mu &\int_{-\infty}^{+\infty} dx \phi(x) e^{i(n_j k_j/Q_0)x - \tau_\mu \Phi(x)} \\
&\times \int_x^{+\infty} dx' \phi(x') e^{-i(n_j k_j/Q_0)x'}.
\end{aligned}
\end{equation}

Notice the appearance of the perturbed line-source function in Eq.~\eqref{eq:Perturbradfield_freqaverage} that still remains from Eq.~\eqref{eq:Transfereq_perturbedstart}. For the perturbed line-source function we follow the procedure in Eq.~\eqref{eq:line_source_func_expression} and from Eq.~\eqref{eq:Perturbradfield_freqaverage} we obtain
\begin{equation}
\delta S_L =  \frac{I_\star}{1-\langle \xi \rangle} \left\langle \left( \mathcal{D}_\mu - \frac{\langle \mathcal{D}_\mu p_\mu \rangle}{\langle p_\mu \rangle}  \right) \frac{\delta Q}{Q_0} \zeta \right\rangle.
\end{equation}

The perturbed Eddington line-flux becomes
\begin{equation}
\delta \bar{H}_i = I_\star \left\langle \left( n_i+ i  \eta_i \right) \left( \mathcal{D}_\mu -  \frac{\langle \mathcal{D}_\mu p_\mu \rangle}{\langle p_\mu \rangle}  \right)\frac{\delta Q}{Q_0}\zeta  \right\rangle.
\end{equation}

Following \citetalias{1985ApJ...299..265O}, in the latter expression we have used the fact that $\mathrm{Re}[\xi] = \mathrm{Re}[-\xi]$ such that $\langle \xi \rangle$ is always real, and that $\mathrm{Im}[\xi] \neq \mathrm{Im}[-\xi]$ making that $\langle n_i \xi \rangle$ is always imaginary. This allows us to rewrite the appearance of $\xi$ into a new quantity
\begin{equation}
\eta_i \equiv \frac{ \langle n_i \mathrm{Im}[ \xi ] \rangle}{1-\langle \mathrm{Re}[\xi] \rangle} = -i \frac{\langle n_i \xi \rangle}{1-\langle \xi \rangle}.
\end{equation}

Physically speaking $\eta_i$ contains contributions from gradients in the line-source function, which potentially alters the propagation properties of unstable structures \citep{1996ApJ...462..894O,1999ApJ...510..355O}. It vanishes both in the short-wavelength and long-wavelength limit of perturbations \citepalias{1985ApJ...299..265O} such that it will not affect any growth or damping properties of radiative waves. The latter being the focus of this work, we will neglect it from now on making that
\begin{equation}
\delta \bar{H}_i = \frac{I_\star}{\varv_{\mathrm{th}}} \left\langle i n_i \left( \mathcal{D}_\mu - \frac{\langle \mathcal{D}_\mu p_\mu \rangle}{\langle p_\mu \rangle}  \right)\frac{n_l k_l n_j}{Q_0} \zeta \right\rangle \delta \varv_j,
\end{equation}
where we have inserted Eq.~\eqref{eq:DeltaQ} under a plane wave perturbation to replace the perturbed $Q$-factor. In order to evaluate the perturbed Eddington line-flux the quantity $\zeta$ has to be known. In general $\zeta$ can be solved using Fourier integrals, but it is more convenient \citepalias{1985ApJ...299..265O} to apply the  \emph{exponential shadowing approximation}
\begin{equation}
\zeta \approx \frac{1/\tau_\mu}{ 1 + i n_j k_j/(2x_\mu Q_0) } = \frac{Q_0 / k_L}{ 1 + i n_j k_j/Q_0},
\end{equation}
where $x(\mu)$ is the blue-edge absorption frequency \citepalias{1985ApJ...299..265O}. In the last equality we have made use of the fact that $x_\mu \approx \phi(x_\mu) \tau_\mu/2$ is a slowly varying function of $\mu$ and of order unity \citepalias{1990ApJ...349..274R}. Since $g_0 \propto \bar{H}_0$ it follows that 
\begin{equation}
\frac{\delta g_i}{g_0} = \frac{\delta \bar{H}_i}{\bar{H}_0}.
\end{equation}

As calculated above, under a plane wave perturbation $\delta \bar{H}_i \propto \delta \varv_j$ and hence $\delta g_i \propto \delta \varv_j$ which is the sought relation in Eq.~\eqref{eq:MHDdispersionrelation}. The perturbed line-force to perturbed velocity tensor satisfies
\begin{equation}
\begin{aligned}
\frac{\delta g_i}{\delta \varv_j} = {}  &\frac{g_0}{\varv_\mathrm{th} \langle \mu \mathcal{D}_\mu p_\mu \rangle} \\
&\times  \left\langle i n_i \left( \mathcal{D}_\mu -  \frac{\langle \mathcal{D}_\mu p_\mu \rangle}{\langle p_\mu \rangle}  \right)\frac{n_l k_l n_j}{Q_0} \frac{Q_0 / k_L}{ 1 + i n_l k_l/Q_0} \right\rangle.
\end{aligned}
\end{equation}

This is the radiative acceleration tensor for a scattering-driven flow and is a special case of the general tensor relation of \citetalias{1990ApJ...349..274R} whereby we have only considered pure scattering.


\bsp	
\label{lastpage}
\end{document}